\documentclass[12pt]{article}
\usepackage{amsmath}
\usepackage{graphicx}%
\usepackage{amsfonts}%
\usepackage{amssymb}
\usepackage{psfrag}

\newcommand{\bea}{\begin{eqnarray}}
\newcommand{\eea}{\end{eqnarray}}
\newcommand{\be}{\begin{equation}}
\newcommand{\ee}{\end{equation}}
\newcommand{\vs}[1]{\vspace{#1 mm}}

\renewcommand{\a}{\alpha}
\renewcommand{\b}{\beta}

\renewcommand{\d}{\delta}

\newcommand{\dsl}{\pa \kern-0.5em /}

\newcommand{\pa}{\partial}

\newcommand{\nn}{\nonumber\\}

\newcommand{\eqn}[1]{(\ref{#1})}

\begin{document}
\topmargin 0pt
\oddsidemargin 0mm

\begin{flushright}

USTC-ICTS-07-23\\




\end{flushright}

\vspace{2mm}

\begin{center}

{\Large \bf Intersecting non-SUSY branes and closed string tachyon
condensation}
\vs{6}

{\large J. X. Lu$^a$\footnote{E-mail: jxlu@ustc.edu.cn}, Shibaji
Roy$^b$\footnote{E-mail: shibaji.roy@saha.ac.in}, Zhao-Long
Wang$^a$\footnote{E-mail: zlwang4@mail.ustc.edu.cn} and Rong-Jun
Wu$^a$\footnote{E-mail: rjwu@mail.ustc.edu.cn} }

 \vspace{4mm}

{\em

 $^a$ Interdisciplinary Center for Theoretical Study\\

 University of Science and Technology of China, Hefei, Anhui
 230026, China\\




\vs{4}

 $^b$ Saha Institute of Nuclear Physics,
 1/AF Bidhannagar, Calcutta-700 064, India}

\end{center}


\begin{abstract}
Following \cite{Bai:2006vv} we here consider the supergravity
solutions representing the charged non-supersymmetric $p$-brane (for
$1\leq p \leq 6$) intersecting with chargeless non-supersymmetric
1-brane and 0-brane of type II string theories. We show how these
solutions nicely interpolate between black $p$-branes and the
Kaluza-Klein ``bubble of nothing'' (BON) by continuously varying
some parameters characterizing the solutions from one set of values
to another. By performing a time symmetric general bubble initial
data analysis, we show that the interpolation  implies a possible
transition from black $p$-branes to  KK BON only for $p\leq 4$, as
in these cases there exist locally stable static bubbles under
certain conditions. However, for $p>4$, black branes always decay
into dynamical bubbles. Contrary to what is known in the literature,
we argue that under certain conditions, the mechanism causing this
transition to stable static bubbles can be the closed string tachyon
condensation for all $p\leq 4$.  We also show that the
configurations indeed contain (chargeless) F-strings required for
the closed string tachyons to appear causing the transitions to
occur.

\end{abstract}
\newpage

\section{Introduction}

In the eighties, while discussing the gravitational properties of
certain soliton solutions in five dimensional Kaluza-Klein (KK)
gravity, Gross and Perry \cite{Gross:1983hb} observed that there
exists a two-parameter family of spherically symmetric solutions in
this theory. This solution interpolates between the Schwarzschild
metric and a soliton solution. The Schwarzschild metric has a
singularity masked by a regular horizon while the soliton is
non-singular. In between, the configuration has a naked singularity.
They interpreted the configuration as describing the exterior
geometry of a collapsing star in parameter space, therefore avoiding
the issue of naked singularity appearing in the configuration. But
they were puzzled by the non-uniqueness of the exterior geometry of
a massive object in the KK theory. As a result, the final state of
collapsing matter appears to have two possibilities: a collapsing
star could end up in a black hole described by the Schwarzschild
metric, or it might end up as a soliton, a totally non-singular
metric. The soliton as we know now is a variant of Witten's ``bubble
of nothing'' (BON) \cite{Witten:1981gj}, namely, the static bubble.

A similar family of solutions as we will show
also exist in supergravities, the low
energy string theories. These are the
specific class of intersecting non-susy branes of type II string
theories, namely, the charged non-supersymmetric $p$-brane (for $1\leq
p \leq 6$) intersecting with chargeless non-supersymmetric 1-brane
and 0-brane of type II string theories. These solutions also nicely
interpolate between black $p$-branes and the non-singular static KK
BON by continuously varying some parameters
characterizing the solutions from one set of values to another.
Again in between, the configuration has a naked singularity.
In this paper we will try to understand the physical meaning of this
interpolation.

Recently in a separate development Horowitz \cite{Horowitz:2005vp}
has argued that under certain conditions black strings and black
branes of type II string theories have dramatic new
endpoints\footnote{Generalizing this to charged and uncharged black
holes (i.e., $p = 0$ case) along with other interesting discussions
and potential realistic applications has also been considered
recently in \cite{Green:2006nv}. We thank the anonymous referee for
bringing this reference to our attention.} to Hawking evaporation in
the form of KK BON \cite{Witten:1981gj}. He arrived at this
conclusion by applying the closed string tachyonic instability to
black strings and branes. Closed string tachyons are known to
develop when the fermions in the theory are taken to satisfy
antiperiodic boundary conditions along one\footnote{A circle with
antiperiodic fermions has a one-loop Casimir energy which will cause
it to contract. We will not address this issue here but refer to its
discussion in \cite{Horowitz:2005vp} as well as in
\cite{Ross:2005ms}. For concreteness, we focus in this paper on the
possibility that the circle reaches string scale on or outside the
horizon as in \cite{Horowitz:2005vp}.} of the compact directions and
the size of the circle becomes of the order of string scale
\cite{Scherk:1978ta,Rohm:1983aq}. Adams et. al. \cite{Adams:2005rb}
have argued that when these winding string tachyons are localized
they can trigger a topology changing transformation as a consequence
of the closed string tachyon condensation. The transitions from
black string (branes) to the KK BON, very similar to the
interpolation we mentioned, is an application of this process.

In ref.\cite{Horowitz:2005vp} Horowitz started from the standard
black fundamental string solution \cite{Horowitz:1991cd,Duff:1993ye}
and compactified the string direction. Since the metric in this case
is multiplied appropriately with a harmonic function, the size of
the string wound along the compact direction varies monotonically
from $L$ (where $L$ is the periodicity of the compact direction) to
zero as we move along the radial direction from infinity to the
singular point. So, at some point in between the size of the circle
becomes of the order of string scale and if this occurs on the
horizon, then the closed string tachyon condensation causes the
circle to pinch off and the resulting state is a bubble which in
this case cannot be static but should expand out. Static bubble
appeared while considering a similar transition from black F-string
containing both F and NS5 charge. However, in order to show that
bubble is the end state of this transition, a time symmetric bubble
initial data analysis has been performed (in the context of
two-charge F-string in space-time dimensions $D=6$ and also with the
vanishing dilaton) and it was found that indeed under certain
conditions ($Q/L^2\ll 1$, where $Q$ is the flux associated with the
bubble) the bubble can be stable and so, the black F-string in those
cases can make a transition to classically stable static bubbles. A
similar transition occurs for black $p$-branes as well and was
briefly mentioned in \cite{Horowitz:2005vp}. However, it was
observed that only for $p = 3$ case\footnote{The other possible
cases are M2 and M5 systems or involve more complicated
configurations such as the D1-D5 system \cite{Ross:2005ms}.} the
final bubble could be stable and static. We will discuss these cases
in more detail in this paper and make new observations regarding the
transition from black $p$-branes to stable static bubbles for all $p
\leq 4$.

In this paper we first construct the interpolating solutions. For
this purpose we start from the supergravity solutions representing
the charged non-susy $p$-brane intersecting with chargeless non-susy
1-brane and 0-brane of type II string theories\footnote{Note that
for $p=1$ supergravity solution represents charged non-susy 1-brane
intersecting with chargeless 0-brane.}. These solutions will be
characterized by four (for $p=1$) or five (for $p>1$) independent
parameters. As we have shown in \cite{Bai:2006vv}, here also it can
be checked that when two (for $p=1$) or three (for $p>1$) of the
four or five parameters characterizing the solutions take some
special values, then these intersecting solutions are identical to
the black $p$-brane solutions. We also observed \cite{Bai:2006vv}
that the parameters characterizing the black brane and the related
dynamics are in a different branch of the parameter space from that
describing the brane-antibrane annihilation process or the open
string tachyon condensation \cite{Bai:2005jr,Lu:2004dp}. Although it
would be interesting to understand the dynamics of the closed string
tachyon condensation in this setup, we, however, point out that the
general intersecting solutions (charged non-susy $p$ brane with
chargeless non-susy 1-brane and 0-brane) in fact interpolate between
the black $p$-branes and the KK BON through a series of classical
solutions. To be specific we find that when two (for $p=1$) or three
(for $p>1$) out of the four (for $p=1$) or five (for $p>1$)
independent parameters characterizing the solutions are continuously
varied from one set of values to another the solutions change from
the black $p$-branes to the KK BON. There is no need to take a
so-called double Wick rotation to obtain the KK bubble and it just
corresponds to another point in the parameter space.

Now in order to understand whether the interpolation  means a
possible transition from black $p$-branes to static KK BON, we go
one step further to show that the bubble is locally stable so that
it does not evolve further perturbatively. We do this by performing
a time symmetric general bubble initial data analysis (Note that we
have now non-vanishing dilaton in $D = 10$.). Our analysis indicates
that only for $p\leq 4$, the initial data can lead to locally stable
static bubble under certain conditions (otherwise they lead to
dynamical bubbles). So, only for these cases the interpolation has
the possibility to be interpreted as the transition from black
branes to static KK BON. When $p>4$, the initial data leads either
to locally unstable static bubbles or to dynamical bubbles.
Therefore, the resulting bubbles are in general dynamical. Bubbles
smaller than the locally unstable static bubbles contract while
larger ones expand. Since the black $p$-branes and the corresponding
bubbles can also be related to each other by the so-called
double Wick rotation, their perturbative stability is correlated as
discussed in \cite{Horowitz:2005vp}. This provides an independent
check of the results obtained through the initial data analysis and
we find precise agreement.

Thus it is expected that for $p \leq 4$, there may exist some
perturbative or non-perturbative mechanism which leads to the
transition from black $p$-branes to static KK BON. This in turn can
be understood as the reason for the existence of the interpolating
solutions we constructed. As these are the topology changing
transitions like what happens for the closed string tachyon
condensation we mentioned, one would expect the transition could be
triggered by the closed string tachyon condensation\footnote{The
correlation between the instabilities of black $p$-branes and the
corresponding
bubbles seems to support this as well, as we will see.} discussed in
\cite{Horowitz:2005vp}. We find that for $p \leq 4$, there exists a
stable static bubble when the ratio $Q/L^{7-p}$ is less than certain
maximum value $C_{\rm max}$ and the black $p$-branes can make a
transition to static KK BON  by closed string tachyon condensation.
When $Q/L^{7-p}
> C_{\rm max}$, the $p$-branes will always decay into dynamical
bubbles. However, only for $p=3$, the black D3-brane can decay
directly into stable static bubble through closed string tachyon
condensation. For $p<3$, the black $p$-branes first decay into
unstable static bubble and then under slight perturbation in the
right direction can eventually settle in the stable static bubble.
But for $p=4$, the black D4-branes first decay into dynamical bubble
and then eventually settle in the stable static bubble. So, in all
these cases, the interpolation can be interpreted to be caused by
the closed string tachyon condensation. For this to hold true, we
actually need all the spatial brane directions to be compact and
their respective asymptotical sizes to be on the same order of
magnitude. For $p=5,6$, as there are no stable static bubbles, the
interpolation does not imply a possible transition in parameter
space. In these cases the black branes can only decay to dynamical
bubbles through closed string tachyon condensation.

We would like to emphasize that if closed string tachyon
condensation could be the possible reason for the transition to
occur, either to stable static or dynamical bubble, there must be
fundamental strings present in the supergravity solutions which is
not obvious in our intersecting solutions. In the present context,
the net charge of fundamental strings must be zero since the
resulting bubble doesn't carry any fundamental string charge coupled
to the NSNS 2-form potential and the charge is conserved during the
transition. This gives us hint where the fundamental strings can be
found. We show that the chargeless non-susy 1-branes present in the
intersecting solutions indeed represent the chargeless non-susy
F-strings, i.e., the equal number of fundamental strings and anti
fundamental strings with zero net charge present in the
configuration\footnote{We thank the anonymous referee for suggesting
us to clarify this.}, which when wound along the compact direction
give rise to closed string tachyon causing the transitions to occur.
We remark that the singular region pinching off from the regular
part of spacetime through closed string tachyon condensation can be
viewed as the decoupling in dynamics. This is supported by the
evidence found in \cite{Horowitz:2006mr}.

This paper is organized as follows. In section 2, we present the construction
of the supergravity solutions interpolating between black D$p$-brane and the
KK BON. We give the time symmetric bubble initial data analysis in section 3.
Then we physically interpret the interpolation in terms of the closed string
tachyon condensation in section 4. Section 5 discusses the origin of F-strings
in the intersecting solutions obtained in section 1 which is necessary for the
closed string tachyon condensation to occur. Then we present our conclusion in
section 6.

\section{Interpolating solutions}

In this section we construct the supergravity solutions
interpolating between black D$p$-brane to static KK BON. As
mentioned in the introduction these solutions would be a class of
intersecting solutions in supergravity, namely, the charged non-susy
$p$-brane intersecting with chargeless non-susy 1-brane and 0-brane.
To obtain them we start from the non-isotropic (in
$(p-q)$-directions) charged non-susy $p$-brane solutions given in
\cite{Bai:2006vv}. This solution actually represents charged
non-susy $p$-branes intersecting with chargeless $(p-1)$-brane,
$(p-2)$-brane, \ldots, upto $q$-brane. These solutions were
constructed by first delocalizing a non-susy $q$-brane in transverse
$(p-q)$ directions and then applying T-duality along the delocalized
directions successively \cite{Lu:2005jc}. But for our purpose here
we need the corresponding solutions with $q=0$ and so we write the
eq.(7) of \cite{Bai:2006vv} with $q=0$ as follows,
\bea\label{nonisotropic}
 d s^2 &=& F^{\frac{p+1}{8}} (H{\tilde
{H}})^{\frac{2}{7-p}} \left(\frac{H}{\tilde
H}\right)^{\frac{p\d_1}{8} + \frac{(3-p)\sum_{i=1}^{p}\delta_{i + 1}
}{2(7-p)}} \left(dr^2 + r^2 d\Omega_{8-p}^2\right)\nn & &  +
F^{-\frac{7-p}{8}} \left(\frac{H}{\tilde H}\right)^{\frac{p \d_1}{8}
+ \sum_{i=1}^{p}\frac{\d_{i + 1}}{2}}\left(-dt^2 \right)\nn & & +
F^{-\frac{7-p}{8}} \left(\frac{H}{\tilde H}\right)^{ \frac{(p -
8)\d_1}{8} +\sum_{j = 1}^{p} \frac{\d_{j + 1} }{2}} \sum_{i = 1}^{p}
\left(\frac{H}{\tilde H}\right)^{- 2\d_{i + 1}} (dx^{i})^2\nn e^{2
\phi} &=& F^{\frac{3-p}{2}} \left(\frac{H}{\tilde
H}\right)^{\frac{\d_1}{2}(4-p) - \sum_{i= 1}^{p} 2\d_{i+ 1}}\nn
F_{[8-p]} &=& \hat{Q} {\rm Vol}(\Omega_{8-p}) \eea Here the metric is
given in the Einstein frame. The various functions appearing in the
solution are defined below, \bea\label{functions} F &=&
\left(\frac{H}{\tilde H}\right)^{\a} \cosh^2\theta -
\left(\frac{\tilde H}{H} \right)^{\b} \sinh^2\theta\nn H &=& 1 +
\frac{\omega^{7-p}}{r^{7-p}},\qquad \tilde H = 1 -
\frac{\omega^{7-p}}{r^{7-p}} \eea where $H$ and $\tilde H$ are two
harmonic functions and $\a$, $\b$, $\theta$, $\d_1$, $\d_2$, \ldots,
$\d_{p+1}$ and $\omega$ are $(p+5)$ integration constants and $\hat{Q}$ is
the charge parameter. The parameters here are not all independent,
but they satisfy the following three relations among themselves
\cite{Bai:2006vv}, \bea\label{parameter} & &\a-\b = a\d_1\nn &
&\frac{1}{2} \d_1^2 + \frac{1}{2} \a (\a -a\d_1) + \frac{2\sum_{i >
j=2}^{p+1}\d_i\d_j} {7-p} = \left(1-\sum_{i=2}^{p+1} \d_i^2\right)
\frac{8-p}{7-p}\nn & & \hat{Q} = (7-p) \omega^{7-p} (\a+\b)\sinh2\theta
\eea where $a=-3/2$. Using the above relations we can eliminate
three of the $(p+6)$ parameters mentioned and therefore, the
solution is characterized by $(p+3)$ independent parameters. From
the isometry of the solutions given in \eqn{nonisotropic}, it is
clear that they actually represent charged (since $F_{[8-p]}$ is
non-zero) non-susy $p$-brane intersecting with chargeless (since all
other form-fields are zero) non-susy $(p-1)$-brane, $(p-2)$-brane,
\ldots, upto 0-brane. For our purpose of showing the closed string
tachyon condensation we do not need these general solutions, but
instead we only need the solutions representing charged non-susy
$p$-brane intersecting with chargeless non-susy 1-brane and 0-brane.
So, we make all the directions in the solutions given in
\eqn{nonisotropic} namely, $x^p$, $x^{p-1}$, \ldots, upto $x^2$
isotropic by putting $\d_3=\d_4= \cdots = \d_{p+1} = \d_0$. So the
only directions which remain anisotropic are $x^1$ and $t$. The
solutions \eqn{nonisotropic} then take the form
\bea\label{isotropic} d s^2 &=& F^{\frac{p+1}{8}} (H{\tilde
{H}})^{\frac{2}{7-p}} \left(\frac{H}{\tilde
H}\right)^{\frac{p\d_1}{8} + \frac{2(3-p)\bar\d }{(7-p)}} \left(dr^2
+ r^2 d\Omega_{8-p}^2\right)\nn & &  + F^{-\frac{7-p}{8}}
\left(\frac{H}{\tilde H}\right)^{\frac{p \d_1}{8} +
2\bar\d}\left(-dt^2 \right) + F^{-\frac{7-p}{8}}
\left(\frac{H}{\tilde H}\right)^{ \frac{(p - 8)\d_1}{8} + 2\bar\d -
2\d_2 } (dx^{1})^2 \nn && + F^{-\frac{7-p}{8}} \left(\frac{H}{\tilde
H}\right)^{ \frac{(p - 8)\d_1}{8} + 2\bar\d - 2 \d_0}\sum_{i=2}^p (d
x^i)^2 \nn e^{2 \phi} &=& F^{\frac{3-p}{2}} \left(\frac{H}{\tilde
H}\right)^{\frac{\d_1}{2}(4-p) - 8 \bar\d} \nn F_{[8-p]} &=& \hat{Q} {\rm
Vol}(\Omega_{8-p}) \eea with the parameter relations \eqn{parameter}
now changed as \bea\label{newparameter}
 & &\a-\b = -
\frac{3}{2}\d_1\nn & &\frac{1}{2} \d_1^2 + \frac{1}{2} \a (\a +
\frac{3}{2} \d_1) + \frac{(p - 1)[2\d_2 + (p - 2)\d_0] \d_0} {7-p} =
\left(1- \d_2^2 - (p - 1)\d_0^2\right) \frac{8-p}{7-p}\nn & & \hat{Q} =
(7-p) \omega^{7-p} (\a+\b)\sinh2\theta \eea Note that in
\eqn{isotropic} we have for convenience defined \be\label{bardelta}
{\bar \d} = \frac{1}{4} \sum_{i=1}^p \d_{i+1} = \frac{1}{4} \d_2 +
\frac{(p-1)}{4} \d_0 \ee Also note that the above solution is valid for
$1 \leq p \leq 6$. However, for $p=1$ there is no $\d_0$ and the
solutions are characterized in that case by four independent
parameters namely, $\d_1$, $\d_2$, $\omega$ and $\theta$. On the
other hand, for $p>1$, the solutions are characterized by five
independent parameters namely, $\d_0$, $\d_1$, $\d_2$, $\omega$ and
$\theta$. Also from the metric in \eqn{isotropic} it is clear that
they represent charged non-susy $p$-brane intersecting with
chargeless non-susy 1-brane and 0-brane. The $p$-brane here is
magnetically charged and the corresponding electrically charged
solutions can be obtained by replacing $F_{[8-p]}$ with $F_{[p+2]} =
e^{(3-p)\phi/2}\ast F_{[8-p]}=(1/2)\sinh2\theta\,d(C/F)\wedge dt
\wedge dx^1 \wedge \cdots \wedge dx^{p}$ where $C=(H/\tilde H)^{\a}
- (\tilde H/H)^{\b}$. Now after making a coordinate transformation
from $r$ to $\rho$ as $r = \rho
\left(\frac{1+\sqrt{f}}{2}\right)^{\frac{2}{7-p}}$ where $f = 1 -
\frac{4\omega^{7-p}}{\rho^{7-p}} \equiv 1 -
\frac{\rho_0^{7-p}}{\rho^{7-p}}$, we can rewrite \eqn{isotropic} as,
\bea\label{newequation} d s^2 &=& G^{\frac{p+1}{8}} f^{-
\frac{\alpha (p + 1)}{16} - \frac{p\d_1}{16} - \frac{(3-p)\bar\d
}{(7-p)} + \frac{1}{7 - p}} \left(\frac{d\rho^2}{f} + \rho^2
d\Omega_{8-p}^2\right)\nn & &  + G^{-\frac{7-p}{8}} f^{\frac{\alpha
(7 - p)}{16} - \frac{p \d_1}{16} - \bar\d}\left(-dt^2 \right) +
G^{-\frac{7-p}{8}} f^{ \frac{\alpha (7 - p)}{16} - \frac{(p -
8)\d_1}{16} - \bar\d + \d_2 } (dx^{1})^2 \nn && + G^{-\frac{7-p}{8}}
f^{ \frac{\alpha (7 - p)}{16} - \frac{(p - 8)\d_1}{16} - \bar\d  +
\d_0}\sum_{i=2}^p (d x^i)^2 \nn e^{2 \phi} &=& G^{\frac{3-p}{2}}
f^{-\frac{\alpha (3 - p)}{4} - \frac{\d_1}{4 }(4-p) + 4 \bar\d} \nn
F_{[8-p]} &=& \hat{Q} {\rm Vol}(\Omega_{8-p}) \eea where in
\eqn{newequation} we have defined $G(\rho) = \cosh^2\theta -
f^{\frac{\a+\b}{2}} \sinh^2\theta$. Note that in general the
solution is well defined only for $\rho \ge \rho_0$ and $\a + \b \ge
0$ but for the case of black $p$-branes for which $\a + \b = 2$,
$\rho$ can be extended to $\rho \ge 0$. The parameter relations
remain exactly as in \eqn{newparameter}. We remark here that the
$p$-branes in the above intersecting solutions \eqn{newequation} or
\eqn{isotropic} are actually the non-susy D$p$-branes, but we can
also write the solutions representing charged non-susy F-string
intersecting with chargeless non-susy 0-brane and also charged
non-susy NS5-brane intersecting with chargeless non-susy 1-brane and
0-brane by S-dualizing these solutions for $p=1$ and $p=5$ cases
respectively. The following argument goes through for these
solutions also.

Now it can be easily checked as was also shown in ref.\cite{Bai:2006vv}
that when
\be\label{blackparameter}
\d_1=-\frac{12}{7}, \quad \d_0=\d_2=-\frac{1}{7}
\ee
which implies from \eqn{newparameter} and \eqn{bardelta}
\be\label{blackother}
\a=\frac{16}{7}, \quad \b=-\frac{2}{7},\quad {\rm and} \quad {\bar \d}
= - \frac{p}{28}
\ee
then the solutions \eqn{newequation} reduce to in the string
frame\footnote{Note that the metric in \eqn{newequation} is given in
the Einstein frame.}
\bea\label{blackbrane}
 ds^2 &=&
\bar{H}^{\frac{1}{2}} \left(\frac{d\rho^2}{f} + \rho^2
d\Omega_{8-p}^2\right) +\bar{H}^{-\frac{1}{2}} \left(- f\,dt^2 +
\sum_{i=1}^p (dx^i)^2\right) \nn e^{2\phi} &=&
\bar{H}^{\frac{3-p}{2}}, \qquad F_{[8-p]} = \hat{Q} {\rm
Vol}(\Omega_{8-p}) \eea where
$\bar{H}=1+\sinh^2\theta\rho_0^{7-p}/\rho^{7-p}$ and
$f=1-\rho_0^{7-p}/\rho^{7-p}$. This is the standard black D$p$-brane
solutions \cite{Horowitz:1991cd, Duff:1993ye}.

Now, if on the other hand, we choose
\be\label{bubbleparameter}
\d_0 = -\frac{2}{7}, \quad \d_1 = \frac{4}{7} \quad {\rm and} \quad
\d_2=\frac{5}{7}
\ee
which implies from \eqn{newparameter} and \eqn{bardelta}
\be\label{bubbleother}
\a=\frac{4}{7}, \quad \b=\frac{10}{7} \quad {\rm and} \quad {\bar \d} =
-\frac{2p-7}{28}
\ee
then the solutions \eqn{newequation} reduce to again in the string frame
as,
\bea\label{bubble}
 ds^2 &=&
\bar{H}^{\frac{1}{2}} \left(\frac{d\rho^2}{f} + \rho^2
d\Omega_{8-p}^2\right) +\bar{H}^{-\frac{1}{2}} \left(- dt^2 +
f\,(dx^1)^2 + \sum_{i=2}^p (dx^i)^2\right) \nn e^{2\phi} &=&
\bar{H}^{\frac{3-p}{2}}, \qquad F_{[8-p]} = \hat{Q} {\rm
Vol}(\Omega_{8-p}) \eea These are precisely the KK bubble solutions
which can be obtained from the black $p$-brane solutions by making a
double Wick rotation on the coordinates $x^1$ and $t$. But note that
here our solutions \eqn{newequation} continuously interpolate
between the black D$p$-branes \eqn{blackbrane} and KK-bubbles
\eqn{bubble} through a series of classical solutions and there is no
need to take any Wick rotation. Note that in order to avoid any
conical singularity in \eqn{bubble} at $\rho=\rho_0$ the circle
$x^1$ must have a periodicity of \be\label{bubbleL} L =
(4\pi\rho_0 \cosh\theta)/(7-p). \ee

Thus we see that \eqn{newequation} in fact represents the
interpolating solution from black $p$-brane to the KK BON when the
parameters characterizing the solution vary from one set of values
\eqn{blackparameter} to another \eqn{bubbleparameter}\footnote{Here
we point out that in order to give a physical interpretation of this
interpolating solution as the transition caused by the closed string
tachyon condensation, as we will discuss in section 4, all the brane
directions (not just $x^1$) would have to be compactified with the
asymptotic sizes of the circles associated with various brane
directions of the same order as $L$. Note also that the parameters
$\theta$ and $\rho_0$ change during the transition.}. So, one
might be tempted to think that this continuous change of parameters
can be physically interpreted as the transition from the black
$p$-brane to static KK BON in parameter space. However, in order to
have this possibility, we first need to make sure that the static KK
BON are stable at least perturbatively. For this, we perform a time
symmetric general bubble initial data analysis which we discuss in
the next section.

\section{Initial data analysis}

In order to understand the physical meaning of the interpolation, as
a first step, we must show that the final bubble configuration is
stable so that it does not decay perturbatively further. For this
purpose, we consider the general bubble solution and perform the
initial data analysis. We take the time symmetric initial data as
usual and consider ten dimensional charged bubble solution with non-zero
dilaton. Let us consider the spatial metric to have the form,
\be\label{bubblemetric} ds^2 = \bar H
(\rho)^{-\frac{7-p}{8}}\left(f(\rho) (dx^1)^2 + \sum_{i=2}^p
(dx^i)^2\right) + \bar H (\rho)^{\frac{p+1}{8}}
\left(\frac{d\rho^2}{f(\rho) h(\rho)} + \rho^2
d\Omega_{8-p}^2\right) \ee where $\bar H (\rho)$ and $f(\rho)$ are
as defined before and $h(\rho)$ will be determined from the
constraint equations. Note that the metric is in the Einstein frame.
The other fields are given as, \be\label{otherfields} e^{2\phi} =
\bar H (\rho)^{\frac{3-p}{2}}, \qquad F_{[8-p]} = \hat{Q} {\rm
  Vol}(\Omega_{8-p})
\ee For the above time symmetric initial data, the  only
constraint\footnote{We don't have additional non-trivial constraints
for $F$ and $\phi$.} that needs to be satisfied is
\be\label{constraint} ^9 R = \frac{1}{2}\left(e^{\frac{p-3}{2} \phi}
F^2 + \partial_\rho \phi
\partial^\rho \phi\right)
\ee
Using \eqn{bubblemetric} and \eqn{otherfields}, we solve the constraint
equation \eqn{constraint} to obtain,
\be
h(\rho) = 1 + \frac{\lambda\left(\rho^{7-p} +
    \rho_0^{7-p}\sinh^2\theta\right)}{2(8-p) \rho^{2(7-p)} - (9-p) \rho^{7-p}
  \rho_0^{7-p} + \left[(9-p)\rho^{7-p} - 2 \rho_0^{7-p}\right]\rho_0^{7-p}
  \sinh^2\theta}
\ee where $\lambda$ is an integration constant. This therefore gives
a three-parameter ($\lambda$, $\rho_0$ and $\theta$) family of time
symmetric, asymptotically flat initial data. Note that when we put
$\theta=0$, the charge parameter $\hat Q$ as well as the dilaton
vanishes and in that case $h(\rho)$ becomes \be h(\rho) = 1 +
\frac{\lambda}{2(8-p)\rho^{7-p} - (9-p) \rho_0^{7-p}} \ee For $p=5$,
this matches exactly with eq.(3.13) of \cite{Horowitz:2005vp} as
expected. In order to avoid conical singularity at $\rho = \rho_0$,
the circle $x^1$ must have a periodicity \bea\label{period} L &=&
\frac{4 \pi \rho_0 \cosh\theta}{7-p} \left(1+\frac{\lambda}
{(7-p)\rho_0^{7-p}}\right)^{-\frac{1}{2}} \nn &=& \frac{4\pi
\hat{\rho}_0^{\frac{7 - p}{2}}}{7 - p}\left(\hat{\rho}_0^{7 - p} -
\frac{\hat{Q}^2}{(7 - p)^2 \hat{\rho}_0^{7 - p}}\right)^{- \frac{5 -
p}{2(7 - p)}} \left[1 + \frac{\lambda}{(7 - p) \left(\hat{\rho}_0^{7
- p} - \frac{\hat{Q}^2}{(7 - p)^2 \hat{\rho}_0^{7 -
p}}\right)}\right]^{- 1/2} \eea where in the second equality we have
defined $\hat{\rho}_0^{7-p} = \rho_0^{7-p} \cosh^2\theta$ and used
the charge expression\footnote{Without loss of generality, we assume
from now on $\hat{Q} \geq 0$ i.e., $\theta \geq 0$. Note that the
charge is now a result of the flux on the non-contractible $S^{8 -
p}$ since there is no longer a singularity to act as the source of
the charge. } $\hat{Q}= (7-p) \rho_0^{7-p} \sinh\theta \cosh\theta$
for the purpose of later comparison.  The ADM mass of these bubbles
can be obtained from the metric \eqn{bubblemetric} as,
\be\label{mass} M =
\frac{\Omega_{8-p}}{2\kappa^2}\left[\frac{(5-p)\hat{Q}^2}{2(7-p)^2
\hat{\rho}_0^{7-p}} + \frac{9-p}{2} \hat{\rho}_0^{7-p} -
\frac{(4\pi)^2 \hat{\rho}_0^{7-p}}{2(7-p)
L^2}\left(\hat{\rho}_0^{7-p} - \frac{\hat{Q}^2}{(7-p)^2
\hat{\rho}_0^{7-p}}\right)^{\frac{2}{7-p}}\right] \ee where
$\Omega_n = 2\pi^{(n+1)/2}/\Gamma((n+1)/2)$ is the volume of an
$n$-dimensional unit sphere and $2\kappa^2 = 16\pi G$, with $G$, the
Newton's constant.  The mass has been calculated from the metric
\eqn{bubblemetric} and we have eliminated the unknown integration
constant $\lambda$ in that expression using \eqn{period} as well as
the following relation between $\rho_0$ and $\hat{\rho}_0$, \be
\rho_0^{7-p} = \hat{\rho}_0^{7-p}\left(1-\frac{\hat{Q}^2}{(7-p)^2
    \hat{\rho}_0^{2(7-p)}}\right)\label{relation}
\ee Eq.\eqn{relation} clearly tells us that\footnote{This can also
be seen from the ratio \be \frac{(7 - p)\hat{\rho}_0^{7 -
p}}{\hat{Q}} = \frac{\cosh\theta}{\sinh\theta} \geq 1.\ee  If we
ignore this constraint, the mass $\rightarrow \infty$ as
$\hat{\rho}_0 \rightarrow 0$, indicating the possibility that the
small bubble cannot contract when $\hat{Q} \neq 0$. Indeed if we
consider a 4-parameter family of initial data as we will mention
later on, there will be no such constraint since in that case the
$\hat{Q}$ is independent of $\rho_0$ and $\theta$.}
$\hat{\rho}_0^{7-p} \geq \hat{Q}/(7-p)$. Note that the mass in
\eqn{mass} takes a positive value $M =
\frac{\Omega_{8-p}}{2\kappa^2} \hat{Q}$ when $\hat{\rho}_0$ takes
the lowest value while  $M \to -\infty$ when $\hat{\rho}_0 \to
\infty$. So, there is no lower bound on mass and the positive energy
theorem fails for our initial data as was also found for the case
studied in \cite{Horowitz:2005vp}.

Thus, for fixed values of $\hat{Q}$ and $L$, the expression for ADM
mass \eqn{mass} has extremum when $dM/d\hat{\rho}_0 = 0$ and this
implies, \be\label{extremeperiod} L = \frac{4\pi
\hat{\rho}_0^{\frac{7-p}{2}}}{7-p}\left(\hat{\rho}_0^{7-p} -
\frac{\hat{Q}^2}{(7-p)^2
\hat{\rho}_0^{7-p}}\right)^{-\frac{5-p}{2(7-p)}} \ee In comparison
with the second expression in
\eqn{period}, we find that the extremum occurs at $\lambda=0$ (or,
$h(\rho)=1$). The resulting metric in \eqn{bubblemetric} is now the
spatial part of the static bubble one obtained from the double
Wick rotations of black $p$-brane in Einstein frame.

    We now come to examine whether the extremum is a local minimum or
a maximum. This will give us indications about the nature of
stability of the static bubbles under consideration. For this, let
us consider $\hat{Q} = 0$ and $\hat{Q} \neq 0$ separately. When
$\hat{Q} = 0$, i.e., $\theta = 0$,  we have from
\eqn{extremeperiod} \be \label{zeroqsolution} \hat{\rho}_0 = \rho_0
= \frac{7 - p}{4\pi} L.\ee In other words, the static bubble size is
always less than $L$, the circle size at infinity.  One can easily
check that the extremum is now a maximum, as expected, by
calculating $ \frac{d^2 M}{d\hat{\rho}^2_0} < 0$ at this
$\hat{\rho}_0$. So this is an unstable static bubble. This also
implies the usual intuition that bubbles which are smaller than the
size of the circle at infinity contract while larger ones expand as
discussed in \cite{Horowitz:2005vp}. Adding flux (i.e., for
non-vanishing charge $\hat{Q} \neq 0$) on the sphere causes it to
expand and the vacuum bubbles which would normally contract could
become stable static bubbles with this flux. So let us turn our
attention to this case next.

For this, let us first determine the number of solutions for
$\hat{\rho}_0$ from \eqn{extremeperiod} for fixed values of $\hat{Q}
\neq  0$ and $L$. This number is actually the same as that for
$\theta$ for a given ratio of $ C = \hat{Q}/L^{7 - p}$ determined
from the following equation \be\label{thetaSL} \frac{\hat{Q}}{L^{7 -
p}} = \frac{(7 - p)^{8 - p}}{(4\pi)^{7 - p}} \frac{\sinh
\theta}{\cosh^{6 - p}\theta} = C (\theta). \ee In obtaining
\eqn{thetaSL}, we have used the expression $\hat{Q} = (7-p)
\rho_0^{7-p} \sinh\theta \cosh\theta$ and the expression
\eqn{bubbleL} for $L$ for the static bubble which holds also for the
extremum as shown above. Once $\theta$ is determined, the $\rho_0$
can be determined either from the $\hat{Q}$ expression or from the
equation \eqn{bubbleL} for $L$. $\hat{\rho}_0$ is then determined
from $\rho_0$. This provides an alternative (also much simpler) way
to determine the number of solutions for $\hat{\rho}_0$ from
\eqn{extremeperiod}. In the following, we consider $p < 5$ cases
first and then the remaining $p = 5, 6$ cases can be considered in a
much direct and simpler way afterwards.

  It is not difficult to check  for $p < 5$ that $C (\theta)$ in \eqn{thetaSL}
has a unique maximum at $\sinh^2 \theta = 1/(5 - p)$. This can also
be seen from the fact that $C (\theta) > 0$ in general and $C
(\theta) \rightarrow 0$ for both $\theta \rightarrow 0$ and $\theta
\rightarrow \infty$. This maximum is given as, \be \label{ratioCS}
C_{\rm max} = \left(\frac{\hat{Q}}{L^{7 - p}}\right)_{\rm max} =
\frac{(7 - p)^{8 - p}(5 - p)^{\frac{5 - p}{2}}}{(4\pi)^{7 - p} (6 -
p)^{\frac{6 - p}{2}}} < 1.\ee So for any given positive ratio $
\hat{Q} /L^{7 - p} = C (\theta) < C_{\rm max}$, we will have two
solutions for $\theta$ from equation \eqn{thetaSL} since the
constant line $C (\theta) = \hat{Q}/{L^{7 - p}}$ will intercept
twice with the curve \be\label{curve} C(\theta) = \frac{(7 - p)^{8 -
p}}{(4\pi)^{7 - p}} \frac{\sinh \theta}{\cosh^{6 - p}\theta}.\ee The
various features mentioned here are shown in Fig.1.

\begin{figure}
 \psfrag{C}{$C(\theta)$}
 \psfrag{Cmax}{$C_{{\rm max}}$}
 \psfrag{thetas}{$\theta_s$}
 \psfrag{theta0}{$\theta_0$}
 \psfrag{thetal}{$\theta_l$}
 \psfrag{theta}{$\theta$}
\begin{center}
\includegraphics{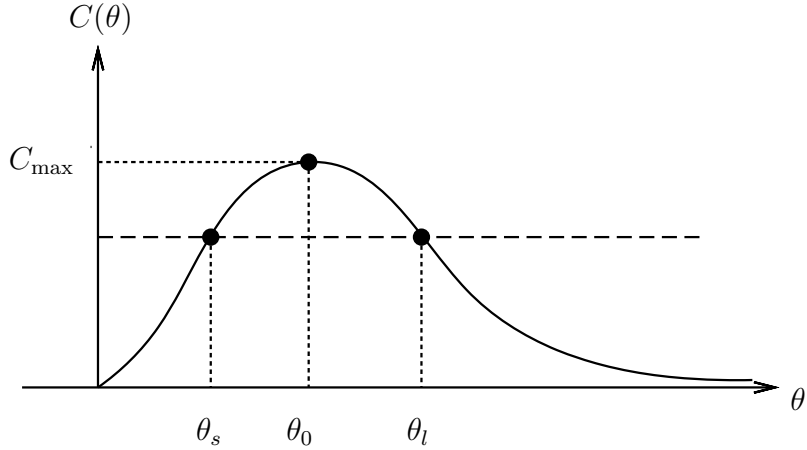}
\end{center}
\caption{The $\theta_0$ gives rise to the maxima of $C (\theta)$,
i.e., $C_{\rm max}$ while $\theta_s$ and $\theta_l$ are two
solutions of \eqn{thetaSL} when $\hat{Q}/L^{7 - p} < C_{\rm max}$
for $p\leq4$} \label{pleq4}
\end{figure}

 Now we can relate
$\hat{\rho}_0$ with $\theta$ by using \eqn{bubbleL} and the relation
$\hat{\rho}_0^{7-p} = \rho_0^{7-p} \cosh^2\theta$ as,
\be\label{rhotheta} \hat{\rho}_0 = \frac{(7-p) L}{4\pi}
\cosh^{\frac{p-5}{7-p}}\theta \ee So, this implies that we have two
solutions for $\hat{\rho}_0$ from \eqn{extremeperiod} under the same
condition, the small $\theta$ corresponds to large $\hat{\rho}_0$
and large $\theta$ corresponds to small $\hat{\rho}_0$. Since these
two solutions are extrema of ADM mass, therefore one extremum must
give a maximum and the other gives the minimum.  Given the fact that
$M \rightarrow - \infty$ as $\hat{\rho}_0 \rightarrow \infty$, the
extremum with the large $\hat{\rho}_0$ must give the maximum while
the one with the small $\hat{\rho}_0$ gives a local minimum. This is
entirely consistent with our anticipation from the $\hat{Q} = 0$
discussion given above. This characteristic feature of the ADM mass
versus $\hat{\rho}_0$ is shown in Fig.2 for $p \leq 4$. Note that,
however, even the local stable static bubble is non-perturbatively
unstable since the ADM mass is unbounded from below.

\begin{figure}
 \psfrag{M}{$M(\hat{\rho}_0)$}
 \psfrag{A}{$\hat{\rho}_0$}
\begin{center}
\includegraphics{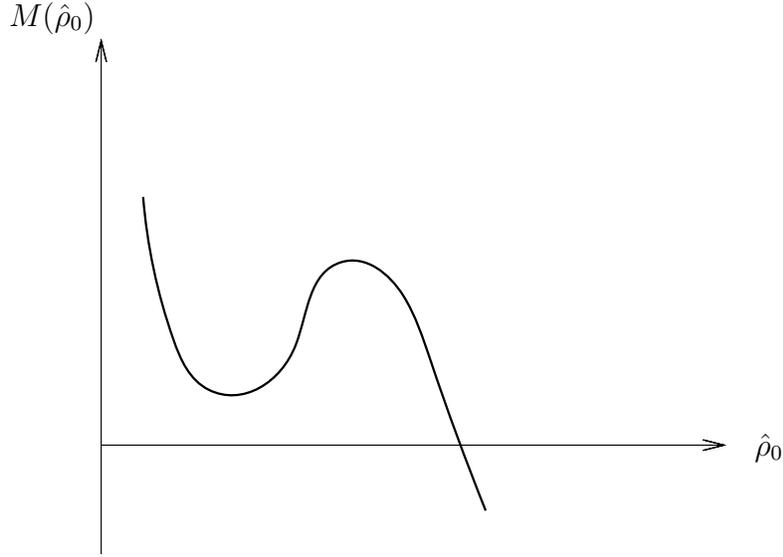}
\end{center}
\caption{The characteristic behavior of the mass vs. $\hat{\rho}_0$
as given in \eqn{mass} for $p \leq 4$. The local minimum is at the
$\hat{\rho}_0$ as given by \eqn{rhotheta} with $\theta = \theta_l$
while the local maximum is at the $\hat{\rho}_0$ determined by the
same equation with now $\theta = \theta_s$.}\label{admass}
\end{figure}

 When the
ratio $\hat{Q}/L^{7 - p}$ takes the maximum value as given in
\eqn{ratioCS}, we expect that the two merge into a single extremum
with $ \frac{d^2 M}{d\hat{\rho}^2_0} = 0 $ when the $\hat{\rho}_0$
takes its extremum value. This can be checked independently from the
following \bea \label{maxmin}\left.\frac{d^2 M}{d
\hat{\rho}_0^2}\right|_{\hat{\rho}_0 = c_0} &=& - \frac{\Omega_{8 -
p}}{2\kappa^2} \frac{9 - p}{7 - p} \frac{(4\pi)^2 c_0^{7 - p}}{L^2}
\left(1 - \frac{\hat{Q}^2}{(7 - p)^2 c_0^{2(7 - p)}}\right)^{-
\frac{2(6 - p)}{7 - p}}\nn &\,& \times \left[1 - (6 -
p)\frac{\hat{Q}^2}{(7 - p)^2 c_0^{2(7 - p)}}\right]\left(1 - \frac{5
- p}{9 - p}\frac{\hat{Q}^2}{(7 - p)^2 c_0^{2(7 - p)}}\right) \eea
where we have set $\hat{\rho}_0 = c_0$ with $c_0$ the solution of
the equation \eqn{extremeperiod}. By demanding $ \left.\frac{d^2
M}{d\hat{\rho}^2_0}\right|_{\hat{\rho}_0 = c_0} = 0 $, we indeed
have one unique solution $c_0^{7 - p} = \sqrt{6- p}\, \hat{Q}/(7 -
p)$ which corresponds to the maximum ratio of $\hat{Q}/L^{7 - p}$ as
anticipated above.

Now we try to find some concrete solutions from \eqn{extremeperiod} under
certain special conditions to support the above general analysis. For this,
we rewrite \eqn{extremeperiod} in the following form,
\be\label{rewriteperiod} \frac{(4\pi)^2 \hat{\rho}_0^2}{(7-p)^2 L^2}
= \left(1-\frac{\hat{Q}^2}{(7-p)^2
\hat{\rho}_0^{2(7-p)}}\right)^{\frac{5-p}{7-p}}. \ee For  $p \leq
4$,  we find the following solutions from \eqn{rewriteperiod} as
\bea (i) \quad \hat{\rho}_0 &\approx& \frac{(7-p) L}{4 \pi},
\qquad\quad {\rm when} \qquad \hat{\rho}_0 \gg
\left(\frac{\hat{Q}}{7-p} \right)^{\frac{1}{7-p}}\qquad \nn (ii)
\quad \hat{\rho}_0 &\approx &
\left(\frac{\hat{Q}}{7-p}\right)^{\frac{1}{7-p}}, \qquad {\rm when}
\qquad \hat{\rho}_0 \ll L \eea So, for case $(i)$, the extremum
occurs at $\hat{\rho}_0 \approx (7-p)L/(4\pi)$, and for case $(ii)$,
the extremum occurs at $\hat{\rho}_0^{7-p} \approx \hat{Q}/(7-p)$.
Note that in both cases $L^{7-p} \gg \hat{Q}$, satisfying the
allowed constraint $L^{7 -p} > \hat{Q}$ in \eqn{ratioCS}.

We thus find that for $L^{7-p} \gg \hat{Q}$, we have two static
bubbles, one large bubble at $\hat{\rho}_0 \approx (7-p)L/(4\pi)$
and one small bubble at $\hat{\rho}_0^{7-p} \approx \hat{Q}/(7-p)$.
 We calculate $d^2M/d\hat{\rho}_0^2$ using \eqn{maxmin} at the two
values of $\hat{\rho}_0$ and find that the large bubble corresponds
to a maximum and the small bubble correspond to a minimum. This is
entirely consistent with our above general analysis.

Now let us consider $p=5,6$. It is clear from the equation
\eqn{rewriteperiod} that in both cases there is only one extremum.
For $p = 5$, the extremum is at $\hat{\rho}_0 = L/(2\pi)$ while for
$p = 6$ it is at $ \hat{\rho}_0 =  \sqrt{L^2/(4\pi)^2 + \hat{Q}^2}$.
Therefore, considering the bound from \eqn{thetaSL}, i.e., $L^2 \geq
2\pi^2 \hat{Q}$, the extremum for $p=5$ gives $\hat{\rho}_0^2 =
L^2/(2\pi)^2 \geq \hat{Q}/2$ which satisfies the constraint
$\hat{\rho}_0^{7-p} \geq \hat{Q}/(7-p)$. For $p = 6$, we don't have
any bound from \eqn{thetaSL} but now the extremum  $ \hat{\rho}_0 =
\sqrt{L^2/(4\pi)^2 + \hat{Q}^2}$ automatically satisfies the
constraint $\hat{\rho}_0^{7-p} \geq \hat{Q}/(7-p)$. The existence of
a single extremum is also consistent with the fact that there exists
a single solution of $\theta$ in \eqn{thetaSL} for either $p = 5$ or
$p =6$. This can be understood from that for $p = 5$, the
$C(\theta)$ has a maximum at $\theta \rightarrow \infty$ while for
$p = 6$, $C(\theta)$ increases monotonically to infinity. The
corresponding features are shown in Fig.3 for $ p = 5$ and in Fig.4
for $p = 6$, respectively.
The extremum for each case is actually a maximum which can be seen
from the sign of the double derivative $d^2M/d\hat{\rho}_0^2$ in
\eqn{maxmin} at the corresponding extremum value of $\hat{\rho}_0$.
In other words, the bubbles in both cases are static but locally
unstable. Note that adding flux on the sphere for $p=5, 6$
does not change the qualitative picture from vanishing flux case.

\begin{figure}
 \psfrag{C}{$C(\theta)$}
 \psfrag{Cmax}{$C_{{\rm max}}$}
 \psfrag{theta}{$\theta$}
\begin{center}
\includegraphics{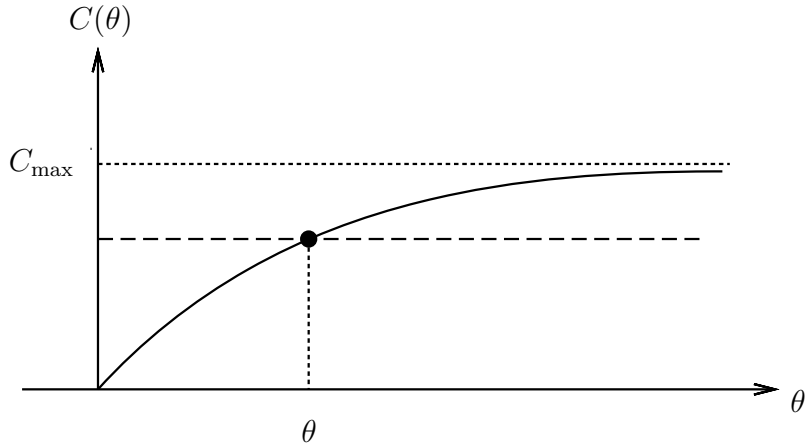}
\end{center}
\caption{The black dot represents a solution of \eqn{thetaSL} with
fixed ratio $\hat{Q}/L^2 < 1/ (2\pi^2)$ for $p=5$. The corresponding
$\theta$ gives a static but unstable bubble.} \label{pe5}
\end{figure}

\begin{figure}
 \psfrag{C}{$C(\theta)$}
 \psfrag{theta}{$\theta$}
\begin{center}
\includegraphics{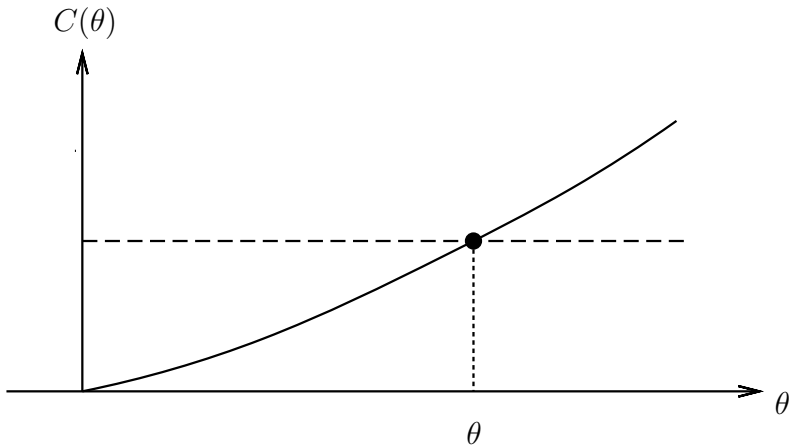}
\end{center}
\caption{The black dot represents a solution of \eqn{thetaSL} with
any given ratio $\hat{Q}/L$ for $p=6$. The corresponding $\theta$
gives again a static but unstable bubble.} \label{pe6}
\end{figure}

So, the general bubble initial data analysis tells us that for $p
\leq 4$, there are two static bubbles if $\hat{Q}/L^{7-p} < C_{\rm max}$. These
bubbles correspond to the doubly Wick rotated form of the black
$p$-branes and of them the smaller bubble is locally stable and the
larger one is locally unstable. On the other hand  for $p = 5, 6$,
there is an unstable static bubble but for $p = 5$  an additional condition
$L^2 > 2\pi^2 \hat{Q}$ needs to be satisfied.

From this analysis, since we found that only for $p \leq 4$ and for
$\hat{Q}/L^{7-p} < C_{\rm max}$, there can be a static  (locally)
stable bubble, therefore, only in those cases the interpolating
solution \eqn{newequation} have the possibility to be interpreted as
the transition from black D$p$-brane to static KK BON. For all other
cases, the black D$p$-branes would decay into dynamical bubbles. We
also performed a 4-parameter family of time-symmetric,
asymptotically flat initial data analysis in which the charge
parameter ${\hat Q}$ is treated to be independent of $\rho_0$ and
$\theta$ and the same conclusion is reached. Therefore, the above
picture is believed to be quite generic. In the following, we will
provide another independent check of the nature of stability of the
static charged bubbles we obtained.

For this, let us recall the discussion  given in
\cite{Horowitz:2005vp} by Horowitz following \cite{Sarbach:2004rm}.
It has been argued there that for black strings and the
corresponding static bubbles, the perturbative stability of bubbles
is directly related to that of the black strings since the two are
related to each other by double Wick rotation. The study of
perturbations of black strings and $p$-branes started with the work
of Gregory and Laflamme \cite{Gregory:1994bj}. It is known that the
short wavelength modes are always stable but the long wavelength
modes are sometimes unstable. The instability of long wavelength
modes further implies a static perturbation and the existence of a
static perturbation  in turn usually indicates an instability. The
Gubser-Mitra conjecture \cite{Gubser:2000ec} states that charged
black strings will be unstable if and only if their specific heat is
negative. Based on the above, Horowitz concluded that for spacetime
dimensions $D = 5, 6$ and large $L$, there is always a static
perturbation of the black string, therefore an unstable perturbation
of the corresponding bubble. In higher dimensions, the static
bubbles are perturbatively stable or unstable depending on how close
to extremality the corresponding black string is.

Since black $p$-branes (for $p \leq 6$) in $D = 10$ are related to black
strings in $D = 11 - p$ through the so-called double dimensional
reduction (here dimensional reductions are performed simultaneously
on both spacetime and brane spatial directions), for example, black
6-branes and 5-branes in D = 10 are related to black strings in D =
5 and 6 respectively, so from the above discussion, we must have black
6-branes and 5-branes in D = 10 unstable and therefore the
corresponding static bubbles are also unstable. On the other hand,
since black $p$-branes with $p \leq 4$ are related to
black strings in $D = 11 - p > 6$, so the perturbative stability of black
$p$-branes with $p \leq 4$, should depend on
how close they are to their respective extremality and this also determines
the stability of the corresponding bubbles. We can understand the
above from the signs of the specific heat of the black
$p$-branes as follows. The specific heat can be calculated once a
black $p$-brane configuration (such as \eqn{blackbrane}) is given, for
example, see \cite{Cai:1997cv}, and its sign is determined by the
sign of the following quantity
\be\label{sign}
\rho_0^{7 - p}\left[(6 - p) \sinh^2\theta - \cosh^2\theta\right].
\ee
It is easy to
see that for $p = 5, 6$, its sign is always negative, signalling the
instability as described above. For $p \leq 4$, its sign indeed
depends on how close the black $p$-branes are to the extremality
which corresponds to taking
$\theta \rightarrow \infty$. As will be discussed in the following
section, the closed string tachyon condensation occurring on or
outside the horizon always requires large $\theta$ for large $L$, so
to leading order \eqn{sign} can be written as,
\be  \frac{(5 - p)}{4} \rho_0^{7 - p}
e^{2\theta} > 0 \ee for $p \leq 4$, and therefore the corresponding
static bubbles are classically stable. This completely independent
analysis supports the results of our initial data analysis.

\section{Physical interpretation of the interpolation}

Let us next try to understand the underlying mechanism for the
transition. The existence of the interpolation indicates a possible
transition from black D$p$-brane to the static and locally stable
KK BON  for $p \leq 4$ (the interpolation makes sense only for these
cases). The closed string tachyon condensation could be the possible
mechanism for these cases if the curvature on the horizon (where the
tachyon condensation takes place) is much smaller than the string
scale, otherwise we will reach the correspondence point where there
is no brane description and the black D-brane would make transition
to the open string modes \cite{Horowitz:2005vp}. For $p>4$, we have
seen that there exist static bubbles, but the bubbles in these cases
are unstable and the corresponding interpolation cannot hold even
true classically since the static bubble will contract or expand
under any small perturbation even though the interpolating solutions
exist. We expect that the closed string tachyon condensation in
these cases if possible must give rise to the corresponding
dynamical bubbles as we will see.

Now from the metric given in \eqn{blackbrane} we see that if the
closed string tachyon condensation occur on the horizon, then the
size of the circle associated with the coordinate $x^1$ is
\be\label{black} L= l_s\cosh^{\frac{1}{2}}\theta, \ee where $l_s$ is
the fundamental string length. Also note that since $L\gg l_s$, the
angle $\theta$ must be very large\footnote{Therefore, the black
$p$-branes must be nearly extremal which guarantees the perturbative
stability of static bubbles for $p \leq 4$ as mentioned in the
previous section.} and therefore $L$ can be approximated as
\be\label{blackL} L = l_s \cosh^{\frac{1}{2}}\theta \approx
\frac{l_s}{\sqrt{2}} e^{\theta/2} \ee The charge of the black
D$p$-branes can also be obtained from the form-field given in
\eqn{blackbrane} as \be \label{blackQ} \hat{Q} = (7-p)\rho_0^{7-p}
\sinh\theta \cosh\theta \approx \frac{7-p}{4}\rho_0^{7-p}
e^{2\theta}\ee Also the size of the horizon can be obtained from the
metric in \eqn{blackbrane} as, \be\label{horizonsize} Z = \rho_0
\cosh^{\frac{1}{2}}\theta \approx \frac{\rho_0}{\sqrt{2}}
e^{\theta/2} \ee Since the curvature at the horizon has to be much
smaller compared to the string scale, so we get,
\be\label{curvature1} {\cal R} \sim \frac{1}{Z^2} \ll
\frac{1}{l^2_s} \quad \Rightarrow \quad \frac{\rho_0}{\sqrt{2} l_s}
\gg e^{-\theta/2} \ee Now we assume that the underlying mechanism
for the transition from black $p$-branes to bubbles is via closed
string tachyon condensation, the final bubbles can be either (stable
or unstable) static or dynamical.  From our initial data analysis in
section 3, we know that we can have stable static bubbles only for
$p \leq 4$ and for $p > 4$ there exist only unstable static bubbles.
So, we discuss the possible transitions for $ p \leq 4$ and $p > 4$
separately. Also for simplicity we will assume that the closed
string tachyon condensation occurs on the horizon. As will be
discussed that there exist situations for which the parameters
characterizing the bubbles can be different from those of the black
$p$-branes, so we denote the bubble parameters with a `tilde'. From
\eqn{bubble} we give below the flux $(Q_b$) associated with the
bubble, the size ($Z_b$) of the bubble and the circle size ($L_b$,
see eq.\eqn{bubbleL} where we rename $L$ as $L_b$) of the circle at
infinity as, \bea\label{bubblequant} Q_b & = & (7-p)
\tilde{\rho}_0^{7-p} \sinh\tilde{\theta} \cosh\tilde{\theta}\nn Z_b
&=& \tilde{\rho}_0 \cosh^{\frac{1}{2}}\tilde{\theta}\nn L_b  &=&
\frac{4\pi}{7-p} \tilde{\rho}_0\cosh\tilde{\theta} \eea We have seen
in section 3 that there exist two static bubbles with small one
stable and the large one unstable for $p \leq 4$ and these two
static bubbles occur at two values of $\tilde\theta$ corresponding
to the two points where the constant line
$C(\tilde\theta)=\hat{Q}_b/L_b^{7-p}$ meets the curve \eqn{curve}.
The curve \eqn{curve} has a maximum at $\sinh\tilde{\theta}_0 =
1/\sqrt{5-p} < 1$ and so, $\tilde{\theta}_0<1$. The two values of
$\tilde{\theta}$ at which the static bubbles exist may be denoted as
$\tilde{\theta}_s$ and $\tilde{\theta}_\ell$ where $\tilde{\theta}_s
< \tilde{\theta}_0 < \tilde{\theta}_\ell$. Let us first take
$\tilde{\theta}_s$ as the static bubble parameter, then since this
is a very small angle $\tilde{\theta}_s < \tilde{\theta}_0 <1$,
\eqn{bubblequant} can be approximated, to leading order, as
\bea\label{bubblequant1} Q_b &= & (7-p) \tilde{\rho}_0^{7-p}
\tilde{\theta}_s\nn Z_b &=& \tilde{\rho}_0 \nn L_b  &=&
\frac{4\pi}{7-p} \tilde{\rho}_0 \eea If the black brane makes a
transition to a static bubble through closed string tachyon
condensation, the charge must be conserved. Also the horizon size of
the black brane must be equal to the bubble size and the size of the
compact circle at infinity must be the same since the local closed
string tachyon condensation should not have much effects on them.

At this point we would like to clarify why we don't identify the
other brane directions (other than $x^1$) as well as the dilaton
profile before and after the closed string tachyon condensation,
i.e., for the black $p$-brane and the BON. It is clear from
\eqn{blackbrane} and \eqn{bubble} that if we do that then the
harmonic functions would have to be identified and this would imply
that the parameters of the two solutions would have to be identified
and therefore the curvature at the horizon would always be of the
order of string scale and the supergravity description would break
down. In other words, with the identifications of the other
directions along the original black branes with those of the
corresponding bubble, the black branes can not decay into static
BON, but can decay only into dynamical BON. It is true that this
situation can not be avoided as long as the other brane directions
remain non-compact. (Here we remark paranthetically that even for
the case of black D3-brane considered in \cite{Horowitz:2005vp}, the
above comments remain valid and in fact there can not be a
transition from black D3-brane to static BON as claimed otherwise in
\cite{Horowitz:2005vp}.) However, we can get around this problem if
we compactify all the spatial directions of the brane with their
asymptotic sizes having the same order of magnitude as $L$ (say).
Now it is clear that if the closed string tachyon condensation
occurs on the horizon then the sizes of all the brane directions
would be of the order of string length there. So the fundamental
string can wrap around any of the spatial directions to give rise to
closed string tachyon and for this reason we should not identify the
directions along the original brane before and after the closed
string tachyon condensation. However, the various spatial directions
along the brane should still be identified asymptotically where the
tachyon condensation has no effect at all since the tachyon
condensation takes an infinite amount of coordinate time to complete
due to the horizon red-shift factor and its effect takes an infinite
amount of coordinate time to reach this region\footnote{Although it
is known \cite{Adams:2005rb} that the local proper time for the
completion of the closed string tachyon condensation on the horizon
is of the order of string scale, however, precisely for this reason,
the tachyon condensation process viewed by an observer at infinity
(with respect to whom the ADM mass is measured) is a very slow
process and therefore to this observer the tachyon condensation can
be taken as an adiabatic process. This is consistent with the
initial data analysis performed in the previous section since it is
based on the ADM mass which is measured by the asymptotic observer
and also with our interpolating solution as we will explain.}. This
is precisely the reason that asymptotically the two configurations
\eqn{blackbrane} and \eqn{bubble} are identical. As for the
transverse sphere, i.e., the horizon and the size of the bubble,
since they are non-contractible and have to be of large size with
the physical radius much larger than the string scale ($l_s$) for
the supergravity description to remain valid (therefore the string
cannot wrap on this sphere to give rise to tachyon condensation),
we then expect their sizes to remain unaffected before and after the
closed string tachyon condensation occurring along the brane
directions. This is the reason we have identified the size of the
horizon with the bubble size. Given the dilaton as the effective
string coupling and the closed string tachyon condensation as a
perturbative process, the dilaton must remain small for the
occurrence of this process. The transition from black brane to BON
by closed string tachyon condensation is a perturbative process and
the dilaton remains small during the transition. So, the dilaton can
not be thought of as an additional spatial direction (as we do when
we uplift the string theory solution to higher dimensions for strong
coupling, therefore large dimensional size) for which one might
think that it should be identified before an after the tachyon
condensation. Now since we are dealing with two different solutions
(black brane and bubble before and after the tachyon condensation),
there is no reason why the dilaton profile should be identified on
the horizon and this is also consistent with the non-identification
of the brane directions discussed above. For definiteness and for
the purpose of modeling this process, in what follows, we choose one
direction, namely $x^1$, to be the direction along which closed
string tachyon condensation takes place but we need to keep in mind
that the tachyon condensation can occur along any spatial direction
of the brane.

Going by the above arguments, we equate \eqn{blackQ}, \eqn{horizonsize}
and \eqn{blackL} with
the corresponding quantities in \eqn{bubblequant1} to get,
\bea\label{bubblequant2} \tilde{\rho}_0^{7-p} \tilde{\theta}_s &=&
\frac{1}{4} \rho_0^{7-p} e^{2\theta}\nn \tilde{\rho}_0 &=&
\frac{1}{\sqrt{2}} \rho_0 e^{\theta/2}\nn \tilde{\rho}_0 &=&
\frac{4\pi}{\sqrt{2}(7-p)} l_s e^{\theta/2} \eea From
\eqn{bubblequant2} we get, \be\label{condition} \rho_0 =
\frac{4\pi}{7-p} l_s, \qquad \tilde{\theta}_s \sim e^{(p-3)\theta/2}
\ee Note from \eqn{curvature1} that with this $\rho_0$, the
curvature remains small compared to string scale since $\theta$ is
very large. Also since $\tilde{\theta}_s$ is very small, the second
condition in \eqn{condition} is consistent as long as $p<3$. Now
since the small $\tilde{\theta}$ corresponds to large bubble which
is static but unstable, so, for $p<3$, the black branes will make
transitions to unstable static bubble by closed string tachyon
condensation. However, since the bubble is unstable, under slight
perturbation, the bubble can either expand further to infinity or it
can contract to eventually settle down to smaller stable static
bubble. This way, black D1 and D2 branes can indeed make a
transition to stable static bubble by closed string tachyon
condensation and under these conditions the interpolations can have
physical interpretation as the closed string tachyon condensation.

Let us next take $\tilde{\theta}_\ell$ as the static bubble
parameter, where the angle can be large (since we have already
considered the case of small angles and the conditions can be met
only for $p<3$ as we have just seen). Now \eqn{bubblequant} can be
approximated, to leading order, as, \bea\label{bubblequant3} Q_b & =
& \frac{7-p}{4} \tilde{\rho}_0^{7-p} e^{2\tilde{\theta}_\ell}\nn Z_b
&=& \frac{1}{\sqrt{2}} \tilde{\rho}_0 e^{\tilde{\theta}_\ell/2}\nn
L_b &=& \frac{4\pi}{7-p} \tilde{\rho}_0 e^{\tilde{\theta}_\ell} \eea
Now equating \eqn{bubblequant3} with \eqn{blackQ}, \eqn{horizonsize}
and \eqn{blackL} we get, \be\label{condition1} \tilde{\rho}_0 =
\rho_0 e^{\frac{\theta-\tilde{\theta}_\ell}{2}}, \qquad
\tilde{\rho}_0 = \rho_0
e^{\frac{2(\theta-\tilde{\theta}_\ell)}{7-p}}, \qquad \rho_0 =
\frac{(7 - p) l_s}{4\sqrt{2} \pi} e^{-\tilde{\theta}_\ell/2} \ee
Note that the first two relations in \eqn{condition1} can be made
consistent only for $p=3$. The last condition when compared with
\eqn{curvature1} can be consistent only if $\theta \gg
\tilde{\theta}_\ell$. Since $\tilde{\theta}_\ell$ corresponds to
small bubble which is static as well as stable, so, in this case the
black D3-brane makes a transition directly to the static stable
bubble through closed string tachyon condensation and the
interpolation has a physical interpretation.

For $p=4$, it is clear from our above analysis that black D4-brane
can not make a transition either to a stable static bubble or to an
unstable static bubble via closed string tachyon condensation
directly. It can only first make a transition to a dynamical bubble.
Then depending on the size of the dynamical bubble created initially
in comparison with that of the stable or unstable static bubble for
a given ratio $\hat{Q}/L^3 < C_{\rm max}$, we can have the dynamical
bubble to expand to infinity if the size is larger than the size of
the static unstable bubble or otherwise to settle down to the static
stable bubble. For this, let us first try to find out the sizes of
the stable static bubble and unstable static bubble, respectively,
for $p=4$ case. The equation determining $\tilde{\theta}_s$ and
$\tilde{\theta}_l$ for given $\hat{Q}=\hat{Q}_b$ and $L=L_b$ is (see
eq.\eqn{thetaSL}), \bea & & \qquad \frac{\hat{Q}^2}{L^6} =
\frac{3^8}{(4\pi)^6}\frac{\cosh^2\tilde{\theta}-1}{\cosh^4\tilde{\theta}}\nn
& & \Rightarrow \cosh^4\tilde{\theta} - 4 \left(\frac{C_{\rm max}
    L^3}{\hat{Q}}\right)^2 \cosh^2\tilde{\theta} + 4\left(\frac{C_{\rm max}
    L^3}{\hat{Q}}\right)^2 = 0
\eea where we have used $C_{\rm max} = 3^4/[2(4\pi)^3]$ from
\eqn{ratioCS}. The solutions to the above equation are given as,
\bea\label{soln} \cosh\tilde{\theta}_s &=&
\sqrt{2}\left(\frac{C_{\rm max} L^3}{\hat{Q}}\right) \left[1 -
\sqrt{1 - \frac{1}{\left(\frac{C_{\rm max}
    L^3}{\hat{Q}}\right)^2}}\right]^{\frac{1}{2}}\nn
\cosh\tilde{\theta}_\ell &=& \sqrt{2}
\left(\frac{C_{\rm max} L^3}{\hat{Q}}\right)
\left[1 + \sqrt{1 - \frac{1}{\left(\frac{C_{\rm max}
    L^3}{\hat{Q}}\right)^2}}\right]^{\frac{1}{2}}
\eea Now let us try to estimate the value of the ratio $(C_{\rm max}
L^3)/\hat{Q}$. Using the expressions \eqn{blackQ} and \eqn{blackL}
we have \be \frac{\hat{Q}}{L^3} =
\frac{3}{\sqrt{2}}\left(\frac{\rho_0}{l_s}\right)^3 e^{\theta/2} \gg
e^{-3\theta/2} e^{\theta/2} = e^{-\theta} \ee Also we have $C_{\rm
max} = 3^4/[2(4\pi)^3] < 1$. Now since we know that in order to have
static bubbles we must have \be C_{\rm max} > \frac{\hat{Q}}{L^3}
\gg e^{-\theta} \ee The above condition can be rewritten as,
\be\label{cond} 1 < \frac{C_{\rm max} L^3}{\hat{Q}} \ll C_{\rm max}
e^{\theta} \ee Since $\theta$ is a very large angle \eqn{cond} gives
us an estimate of the ratio $(C_{\rm max} L^3)/\hat{Q}$. Now when
the ratio $(C_{\rm max} L^3)/\hat{Q}$ is greater than but close to
1, then from \eqn{soln} we find \be \label{Sol}
\cosh\tilde{\theta}_s \approx \cosh\tilde{\theta}_\ell \approx
\sqrt{2}\left( \frac{C_{\rm max} L^3}{\hat{Q}}\right) \ee and when
$(C_{\rm max} L^3)/\hat{Q} \gg 1$, $\cosh\tilde{\theta}_\ell$ is
still given by almost the same expression as given above in terms of
$L^3/\hat{Q}$, but now $\cosh\tilde{\theta}_s \to 1$ and that
implies $\tilde{\theta}_s \to 0$. So we have from \eqn{bubbleL}, \be
\tilde{\rho}_0 = \frac{3}{4\pi} \frac{L}{\cosh\tilde{\theta}_\ell}
\sim \frac{\hat{Q}}{ L^2} \ee and therefore we have the size of the
stable static bubble as, \be\label{stablebubblesize} Z_b \sim
\tilde{\rho}_0\cosh^{\frac{1}{2}}\tilde{\theta}_\ell \approx
\left(\frac{ \hat{Q}}{L}\right)^{1/2} \ee Now we find out the size
of the dynamical bubble at the moment of its formation through
closed string tachyon condensation as, \be\label{dynamical} Z_{\rm
dyn} = \frac{\rho_0}{\sqrt{2}} e^{\theta/2} \sim
\left(l_s\frac{\hat{Q}}{L}\right)^{1/3} \ee where in the last
expression we have used \eqn{blackQ} and \eqn{blackL}. So, comparing
\eqn{stablebubblesize} and \eqn{dynamical} we get \be\label{Zratio}
\frac{Z_b}{Z_{\rm dyn}} =
\frac{\left(\frac{\hat{Q}}{L}\right)^{1/6}}{l_s^{1/3}} \gg 1 \ee In
the last inequality we have first used \eqn{blackQ} and \eqn{blackL}
to obtain $\rho_0 \approx \hat{Q}^{1/3} (l_s/L)^{4/3}$ and then from
$\rho_0/l_s \gg e^{-\theta/2} \approx l_s/L$ we have $\hat{Q}/L \gg
l_s^2$. Note that eq.\eqn{Zratio} tells us that the size of the
dynamical bubble is much less than the stable static bubble and
therefore, we expect that the dynamical bubble will evolve to
eventually settle down to the stable static bubble. So, the black
D4-brane will first decay to the dynamical bubble by closed string
tachyon condensation and then evolve to settle down to the static
stable bubble plus radiation and this would be the physical
interpretation of the interpolation in this case.

Now let us next look at the cases $p=5,6$. For $p=5$ we have from
\eqn{blackQ}, \eqn{blackL}, \eqn{horizonsize} and \eqn{curvature1}
$\hat{Q}/L^2 = (\rho_0/l_s)^2 e^{\theta} \gg e^{-\theta} e^{\theta} =1$.
On the other hand in order to have static bubble we need (see \eqn{thetaSL})
$\hat{Q}/L^2 = \sinh\tilde{\theta}/(2\pi^2 \cosh\tilde{\theta}) <1$ and so,
black D5-branes can decay only to dynamical bubbles by closed string tachyon
condensation. Similarly for $p=6$, we have $\hat{Q}/L = (\rho_0/l_s)
e^{3\theta/2} \gg e^\theta$. But, in order to have static bubble
we need $\hat{Q}/L \sim \sinh\tilde{\theta} < e^{\tilde{\theta}} \sim
e^{\theta}$ and therefore again we find that the bubble must be dynamical
and expand to infinity.

 In summary, for $p \leq 4$, when the black
$p$-branes (which is compact with the compact directions having the
sizes of the same order of magnitude)  make transitions to the
corresponding bubbles via closed string tachyon condensation, they
can either decay to unstable static bubble and eventually settle
down to the stable static bubble (this happens for
$p<3$)\footnote{The unstable static bubble can also have the
possibility to expand to infinity.} or they can directly decay to
stable static bubble (this happens for $p=3$) or they can decay to
dynamical bubble and then settle down to static stable bubble (this
happens for $p=4$). So, in all these cases of $p \leq 4$, the black
branes can make transitions to static stable bubbles via closed
string tachyon condensation and all the conditions for this to
happen can be met as we have seen. Therefore, this would be the
physical meaning of the interpolation of our solutions we
constructed in section 2 in parameter space. Note that as we point
out at the beginning of this section for $p = 5, 6$ cases, the
interpolation cannot hold true even classically since the unstable
static bubble will contract or expand under any small perturbation.
The conclusion reached above via closed string condensation is
entirely consistent with this fact and the bubbles formed this way
must be dynamical and expand to infinity. In other words, the
meaningful interpolation of our solutions exists only for $p \leq 4$
and for the branes with their brane directions compact and with all
the compact sizes the same order of magnitude and this can be
explained via the closed string condensation as described above.

Before closing this section, we remark that the black $p$-brane has a
large entropy while the corresponding KK bubble at the moment of its
formation, via closed string tachyon condensation, has no intrinsic
entropy. Therefore, the closed string tachyon condensation must
produce some radiation in addition to the bubble as pointed out in
\cite{Horowitz:2005vp}. This implies that the mass of the bubble
must be less than the mass of the original black $p$-brane. We checked
that this is indeed true.

\section{Origin of F-strings}

We have shown above that the supergravity solutions representing
charged non-susy $p$-brane intersecting with chargeless non-susy
1-brane and 0-brane interpolate continuously between black
D$p$-branes and  KK BON when some parameters characterizing the
solutions are varied from one set of values to another. The
interpolation can be interpreted physically as the transition from
black D$p$-brane to  KK BON only for $p \leq 4$, since only for
these cases there exist static, locally stable bubbles as our
initial data analysis suggests. The underlying mechanisms, as we
have shown, for these transitions could be the closed string tachyon
condensation for all $p \leq 4$.

As we have already noticed the interpolating solution \eqn{newequation}
is well behaved at two end points, i.e. it is regular for the bubble solution
and singular for the black D$p$-brane solution at $\rho=0$ which is masked
by a regular horizon. But for
all other intermediate points the solutions have naked singularities
at $\rho=\rho_0$. Two most natural questions which might
arise at this point are
\begin{itemize}
\item How do we
interpret the intermediate singular solutions?

\item Where are the F-strings in
our set up which when wind along the periodic coordinate $x^1$ give
rise to tachyonic mode when the size of the circle reaches the
string scale on the horizon?

\end{itemize}

To answer the first question we remark here that the intermediate
solutions \eqn{newequation} (with the parameter values other than
those given in \eqn{blackparameter}, \eqn{blackother},
\eqn{bubbleparameter}, \eqn{bubbleother}) are all regular in the
region $\rho_0 < \rho \leq \infty$ and the naked singularity at
$\rho=\rho_0$ reflects our inability to describe the system
classically where the violent quantum process like the closed string
tachyon condensation is occurring. It is very likely that quantum
mechanically there are tachyon condensates \cite{Horowitz:2006mr}
and no singularities, but classically we do not have a good
description in general for the region $\rho \leq \rho_0$.  To an
observer far away from the core region (where the closed string
tachyon condensation is occurring), only the long-range force would
appear and so we have a classical description of the dynamics there
and the description of the tachyon condensation can be viewed as an
adiabatic process as discussed in the previous section. This
long-distance description is just the family of the intermediate
solutions with naked singularities. However, as we stressed, these
singularities are just the artifact since their appearance is due to
the extrapolation of our solutions valid only at long distance to
the region where the description is invalid and where we actually
have quantum process without any singularity. Here our remarks are
very similar in spirit with the family of solutions discussed by
Gross-Perry \cite{Gross:1983hb} and also is supported by the
observations made in \cite{Dine:2006we}.

To answer the second question we point out that although our
intersecting solutions \eqn{isotropic} or \eqn{newequation} does not
seem to contain F-strings to cause the closed string tachyon
condensation, but we will show that \eqn{isotropic} indeed contain
chargeless non-susy F-strings in the sense defined in the
Introduction and this is the reason we have the closed string
tachyon condensation and consequently the transition from
black-branes to KK BON\footnote{There are many ways to see how such
F-strings can appear. For example, the original non-isotropic
solutions (1) contain chargeless $(p - 1)$-brane, $(p - 2)$-brane,
upto 0-brane if we take $q = 0$ there. These chargeless branes are
either brane-antibrane or non-BPS brane in the theory. The process
of making the non-isotropic directions $x^p, x^{p - 1}, \cdots$ upto
$x^2$ isotropic is actually the process of elimination of the
chargeless $(p -1)$-brane, $(p - 2)$-brane upto 2-brane from the
solution as mentioned earlier. This elimination can also be viewed
as the end of annihilation process for these branes. The previous
work \cite{Yi:1999hd, Bergman:2000xf, Lu:2005jc} indicates that the
annihilation of these branes will give rise to F-strings wrapping
around $x^1$ non-perturbatively. In addition, the chargeless part of
non-susy charged D$p$ as well as part of the chargeless  D-strings
in this configuration can also give rise to F-strings when they
annihilate.}. Now in order to show this we will compare
\eqn{isotropic} with another known solutions constructed in
ref.\cite{Lu:2005jc} representing charged non-susy F-strings
intersecting with chargeless non-susy D$p$-branes. The solutions are
given in eqs.(11) and (12) in \cite{Lu:2005jc} and have the forms,
\bea\label{fstring} ds^2 &=& {\bar F}^{\frac{1}{4}} (H{\tilde
{H}})^{\frac{2}{7-p}} \left(\frac{H}{H}\right)^{-(2\sum_{i=2}^p
\gamma_i)/(7-p)}\left(dr^2 + r^2 d\Omega_{8-p}^2\right)\nn && +
{\bar F}^{-\frac{3}{4}}\left(-dt^2 + (dx^1)^2 \right) + {\bar
F}^{\frac{1}{4}}\sum_{i=2}^p \left(\frac{H}{\tilde
H}\right)^{2\gamma_i} (dx^i)^2\nn e^{2\phi} &=& {\bar F}^{-1}
\left(\frac{H}{\tilde {H}}\right)^{2\gamma_1}, \quad
B^{(2)}\,\,=\,\, \frac{\sinh2\theta}{2}\left(\frac{\bar C}{\bar
F}\right)dt \wedge dx^1 \eea with the parameter relation
\be\label{parameterreln} \frac{1}{2}\gamma_1^2 + \frac{1}{2}
\bar\a(\bar\a-\gamma_1) + \frac{2\sum_{i>j=2}^p \gamma_i
\gamma_j}{7-p} = \left(1-\sum_{i=2}^p \gamma_i^2\right)
\frac{8-p}{7-p} \ee Note that here we have labelled the various
functions and the parameters differently to compare the possible
differences with the configuration \eqn{isotropic}. Let us remark
that here the D$p$-branes are chargeless and also we don't have the
presence of chargeless 0-brane but F-strings are charged. So in
making identification, we need to send the F-string charge to zero
which can be done by setting $\theta = 0$. We also need to set
$\gamma_2 =\gamma_3 =\cdots =\gamma_p = \gamma_0$ in order to make
the directions $x^2$, \ldots, $x^p$ isotropic. This way we end up
with solutions representing chargeless non-susy F-strings
intersecting with chargeless non-susy D$p$-branes as follows:
\bea\label{chargelessf} ds^2 &=& (H{\tilde {H}})^{\frac{2}{7-p}}
\left(\frac{H}{H}\right)^{\frac{\bar\a}{4} -\frac{2(p -
1)\gamma_0}{(7-p)}}\left(dr^2 + r^2 d\Omega_{8-p}^2\right)\nn && +
\left(\frac{H}{H}\right)^{-\frac{3\bar\a}{4}}\left(-dt^2 + (dx^1)^2
\right) +  \left(\frac{H}{\tilde H}\right)^{\frac{\bar\a}{4} +
2\gamma_0} \sum_{i=2}^p (dx^i)^2 \nn e^{2\phi} &=&
\left(\frac{H}{\tilde {H}}\right)^{-\bar\a + 2\gamma_1}, \quad \eea
with the parameter relation \be\label{newreln} \frac{1}{2}\gamma_1^2
+ \frac{1}{2} \bar\a(\bar\a-\gamma_1) + \frac{(p - 1)(p -
2)\gamma_0^2}{(7-p)} = \left(1- (p - 1)\gamma_0^2\right)
\frac{8-p}{7-p} \ee Now to compare this with \eqn{isotropic} we must
put the D$p$-brane charge to zero and also to remove the presence of
0-branes we must put $p\d_1/8+ 2{\bar \d} = (p-8)\d_1/8 + 2{\bar \d}
- 2\d_2 \Rightarrow \d_2 = -\d_1/2$. Then we have from
\eqn{isotropic} \bea\label{chargelessd} d s^2 &=&  (H{\tilde
{H}})^{\frac{2}{7-p}} \left(\frac{H}{\tilde
H}\right)^{\a\frac{p+1}{8} + \frac{p\d_1}{8} + \frac{2(3-p)\bar\d
}{(7-p)}} \left(dr^2 + r^2 d\Omega_{8-p}^2\right)\nn & &  +
\left(\frac{H}{\tilde H}\right)^{-\a \frac{7-p}{8}+ \frac{p \d_1}{8}
+ 2\bar\d}\left(-dt^2  + (dx^1)^2\right) + \left(\frac{H}{\tilde
H}\right)^{-\a\frac{7-p}{8} \frac{(p - 8)\d_1}{8} + 2\bar\d - 2
\d_0}\sum_{i=2}^p (dx^i)^2 \nn e^{2 \phi} &=&  \left(\frac{H}{\tilde
H}\right)^{\alpha\frac{3-p}{2} + \frac{\d_1}{2}(4-p) - 8 \bar\d}
\eea and the parameter relation (the second equation of
\eqn{parameter}) takes the form: \be\label{anothernewreln}
\frac{1}{2} \d_1^2 + \frac{1}{2} \a (\a + \frac{3}{2} \d_1) +
\frac{(p - 1)[-\d_1 + (p - 2)\d_0] \d_0} {7-p} = \left(1- \d_1^2/4 -
(p - 1)\d_0^2\right) \frac{8-p}{7-p} \ee If we now try to identify
\eqn{chargelessf} with \eqn{chargelessd}, we find that they can
indeed be identified if we relate the parameters in the two
solutions as follows: \bea\label{identification}
 \bar\a &=& - \frac{2(p - 1)\d_0}{3} +
\frac{(2 - p)\d_1}{6} + \frac{7 - p}{6}\a\nn \gamma_0 &=& \frac{p
- 4}{3} \d_0 + \frac{p - 8}{12} \d_1 - \frac{7 - p}{12}\a\nn
\gamma_1 &=& - \frac{4(p - 1)}{3}\d_0 + \frac{5 - p}{3} \d_1 +
\frac{4 - p}{3} \a.
\eea
Using these relations, we can show that the parameter relation \eqn{newreln}
indeed goes over to \eqn{anothernewreln}. This therefore shows that the
supergravity configuration \eqn{isotropic} or \eqn{newequation} contains
a chargeless F-string causing the tachyon condensation as well as the
transition to occur.

\section{Conclusion}

So, to summarize, in this paper we have explicitly constructed a
class of type II supergravity solutions which interpolate between
black D$p$-brane and KK BON when we vary the parameters
characterizing the solutions from one set of values to another and
obtained a physical meaning of this interpolation. The existence of
such an interpolation in parameter space usually means a transition
from black D$p$-branes to KK BON. For this to be true indeed, the
final bubble configuration must be static as well as stable at least
classically. By performing a time symmetric initial data analysis we
have shown that classically stable, static bubble can result only if
$p \leq 4$ under certain conditions. For all other cases, we found
that the bubbles are either static, but classically unstable or
dynamical and therefore the interpolating solutions in these cases
do not imply a real transition. After we understood this we looked
for the underlying mechanism causing this transition. We found that
except for $p=5,6$, in all other cases the interpolation can be
understood as a real transition from black $p$-branes to the static
KK BON  caused by the closed string tachyon condensation. For $p<3$,
the black branes first make transition to unstable static bubbles
which eventually settle down to the stable static bubbles, for
$p=3$, black D3-brane makes direct transition to the static stable
bubble, whereas for $p=4$, the black brane first makes transition to
the dynamical bubble and then evolves finally to the static stable
bubble. To have the transition to occur actually, we must have all
the brane directions compact with their sizes of the same order of
magnitude and the same must also be true for the corresponding
bubbles. In all these cases the various conditions required for the
closed string tachyon condensation to occur can be satisfied and
therefore the interpolation we found in section 2 can be physically
interpreted as the closed string tachyon condensation. Finally, in
order to claim that our solutions in fact interpolate between black
D$p$-brane (for $p \leq 4$) to the static KK BON or to dynamical
bubbles by closed string tachyon condensation, we must show that the
solutions contain F-strings which is not apparent in our
intersecting solutions. We have explicitly shown this by comparing
with another known solutions that our solutions indeed contain
chargeless F-strings causing the closed string tachyon condensation
which in turn leads the transition to occur.

\section*{Acknowledgements:}

We  wish to thank Hua Bai for collaboration at an early stage of
this work, Sudipta Mukherji for useful discussions and the anonymous
referee for the suggestions and comments which help us to improve
the manuscript. JXL, ZLW and RJW acknowledge support by grants from
the Chinese Academy of Sciences, a grant from 973 Program with grant
No: 2007CB815401 and grants from the NSF of China with Grant
No:10588503 and 10535060.

\end{document}